\documentclass[11pt]{article}
\usepackage{times}
\usepackage{geometry}
\usepackage[utf8]{inputenc}
\usepackage{enumitem,amssymb}
\usepackage{graphicx}
\usepackage{authblk}
\usepackage{url}
\usepackage{soul}
\usepackage[normalem]{ulem}
\usepackage[
 colorlinks=true,    % false: boxed links; true: colored links 
 linkcolor=blue,     % color of internal links            
 citecolor=blue,     % color of links to bibliography
 filecolor=blue,  % color of file links 
 urlcolor=blue,      % color of external link
 final=true
]{hyperref}
\usepackage[numbers,sort&compress]{natbib}
\usepackage{enumitem}
\usepackage[symbol]{footmisc}
\usepackage[dvipsnames]{xcolor}

\setenumerate{itemsep=0mm}

\begin{document}

\title{\large ~~~~~~~~~~~~~~~~~~~~~~~~~~~~~~~~~~~~~~~~~~~~~~~~~~~~~~~~~~~~~~~~~~ Snowmass  2021 KIAS--P22040\\
\LARGE  MSSM Under Higgs Factories}
\author[a]{Honglei Li}
\author[b]{Huayang Song}
\author[c]{Shufang Su}
\author[d]{Wei Su}
\author[b]{Jin Min Yang}

\affil[a]{\small School of Physics and Technology, University of Jinan, Jinan, Shandong 250022, China }
\affil[b]{\small CAS Key Laboratory of Theoretical Physics, Institute of Theoretical Physics,
Chinese Academy of Sciences, Beijing 100190, China.}
\affil[c]{ \small Department of Physics, University of Arizona, Tucson, Arizona 85721, USA}
\affil[d]{\small Korea Institute for Advanced Study, Seoul 02455, Korea}
\maketitle

% \normalsize
\footnotetext[1]{email address: \url{sps\_lihl@ujn.edu.cn,
shufang@arizona.edu,
huayangs@itp.ac.cn,
weisu@kias.re.kr, jmyang@itp.ac.cn}}
\noindent {\large \bf Thematic Areas:}  %(check all that apply $\square$/$\blacksquare$)

\noindent $\blacksquare$ (EF02) EW Physics: Higgs Boson as a portal to new physics \\
\noindent $\blacksquare$ (EF01) EW Physics: Higgs Boson properties and couplings \\
\noindent $\blacksquare$ (EF08) BSM: Model specific explorations \\
\noindent $\blacksquare$ (TF07) Collider phenomenology \\ 
% \noindent {\large \bf Abstract:} (maximum 200 words)
\begin{abstract}
\large 
The high precision measurements of the Higgs mass and couplings at the future Higgs factories are sensitive to the  parameter space of the Minimal Supersymmetric Standard Model (MSSM). Focused on the dominant stop sector contributions, we study the implication of the Higgs precision measurements on MSSM using multi-variable $\chi^2$ fit.  The results show nice complementarity between the indirect searches at Higgs factories and the direct searches at the current LHC program.
\end{abstract}

% \clearpage

\large 

\section{ Introduction}
Direct searches on the  new physics beyond the Standard Model (SM) are extensively explored at the Large Hadron Collider (LHC). Complementary to the direct searches,  the Higgs precision measurements provide an alternative   way to study the new physics effects.  There are several proposals to build Higgs factories in the pursuit of precision Higgs measurements,  including the Circular Electron Positron Collider (CEPC) in China~\cite{CEPCStudyGroup:2018ghi}, the electron-positron stage of the Future Circular Collider (FCC-ee) at CERN~\cite{Mangano:2018mur,Benedikt:2018qee}, and the International Linear Collider (ILC) in Japan~\cite{Bambade:2019fyw}. These future Higgs factories will be the frontier of the precision measurements which will be sensitive to the new physics beyond the SM.

As a well-motivated model to solve problems like the naturalness problem, and the origin of dark matter,  
the MSSM is one of the most promising new physics scenarios. Each SM particle has its supersymmetric partner with spin differed by a half.    The Higgs sector of the MSSM follows that of the Type-II  two Higgs doublet model with one Higgs doublet couples to the up-type quarks, and the other doublet couples to the down-type quarks and charged leptons.  After the electroweak symmetry breaking, there are five physical fields labeled as $A$, $h$, $H$ and $H^{\pm}$, in which $h$ and $H$ are CP-even bosons and $A$ is the CP-odd one.  

\section{MSSM and Analyses Strategy}
The MSSM Higgs sector at the tree level is   described by only  two input parameters $m_A$ and $\tan\beta$,  the mass of the CP-odd Higgs boson and the ratio of the vacuum expectation of the two Higgs fields.   Both the mass and the couplings of the SM-like light CP-even Higgs $h$ receive radiative corrections, which are sensitive to the model parameters.
The CP-even Higgs mass matrix is given by 
\begin{equation}
 \cal{M}_{\rm Higgs} \rm = \frac{ \sin 2 \beta }{2}\left( \begin{array}{ll}
     \cot \beta \ m_Z^2 + \tan \beta \ m_A^2 & ~~~~- m_Z^2 - m_A^2\\
     - m_Z^2 - m_A^2  & \tan \beta \ m_Z^2 + \cot \beta \ m_A^2
     \end{array} \right) +
     \left( \begin{array}{ll}
     \Delta_{11} & \Delta_{12}
     \\
     \Delta_{12} & \Delta_{22}
     \end{array}
     \right),
\label{eq:glalphaap}
\end{equation}
with the first term being the tree-level contributions and   $\Delta_{11},\Delta_{12},\Delta_{22}$ in the second term are the loop-induced Higgs mass corrections~\cite{Dabelstein:1995js,Carena:1995bx,Harlander:2017kuc}. 
 In particular, the dominant contribution to the SM-like Higgs mass comes from the stop sector. To capture the dominant effects, we mainly consider four parameters $\tan\beta, m_A,  m_{\rm SUSY}, X_t$,  with $m_{\rm SUSY}$ being the soft SUSY breaking parameter for stop masses, and $X_t$ being the left-right mixing in the stop mass matrix: $X_t = A_t - \mu  \cot \beta$.  
 In our study, we use the effective mixing angle $\alpha_{eff}$ to calculate the Higgs couplings following the Type-II 2HDM case~\cite{Li:2020glc}, which takes the form of
\begin{eqnarray}\label{eq:alp_eff}
  \tan \alpha_{eff} &=& \frac{-(m_A^2+m_Z^2)\sin\beta \cos \beta + \Delta_{12}}{m_Z^2 \cos^2 \beta +m_A^2 \sin^2 \beta +\Delta_{11}-m_{h^0,eff}^2}.
\end{eqnarray}

To study the sensitivity of MSSM parameters to the Higgs precision measurements at Higgs factories, we  use a multi-variable $\chi^2$ fit:
\begin{equation} 
  \chi^2_{total} = \chi^2_{m_h} + \chi^2_{\mu} = \frac{(m_h^{\rm MSSM}-m_h^{\rm obs})^2}{(\Delta m_h)^2} + \sum_{i=f,V..} \frac{(\mu_i^{\rm MSSM}-\mu_i^{\rm obs})^2}{(\Delta \mu_i)^2},
  \label{eq:chi2}
\end{equation}
here  $\chi^2_{m_h}$ and $\chi^2_{\mu}$ are the contributions to the overall $\chi^2$ from the Higgs mass and signal strength  $\mu_i^{\rm{MSSM}}=(\sigma\times\textrm{Br}_i)_{\rm{MSSM}}/(\sigma\times\textrm{Br}_i)_{\rm{SM}}$  measurements, respectively. For $\chi^2_{m_h}$, given the small experimental uncertainties, we set $\Delta m_h$ to be 3 GeV which mainly comes from    theoretical uncertainty of unknown higher order radiative corrections~\cite{Degrassi:2002fi,Frank:2006yh,Hahn:2013ria,Bahl:2016brp}. In our analyses, we determine the allowed parameter region at the $95\%$ Confidence Level (C.L.) by a multi-variable fit to the Higgs decay signal strengths of various channels and Higgs mass.    

\section{$\chi^2$ Fit Results}

Using the $\chi^2$ analysis, we  study the effects of Higgs precision measurements from   Higgs factories on the MSSM parameter space~\cite{Li:2020glc}.  We study the individual constraint from the Higgs mass, loop induced $h\gamma\gamma+hgg$ channels, as well the channels of Higgs decays to a pair of fermions or gauge bosons.    We found CP-odd Higgs mass $m_A$ is sensitive to the precisions of Higgs decay channels,  while $m_{\rm SUSY}$, $X_t$ and $\tan\beta$ are sensitive to the precision of Higgs mass determination. For large $\tan\beta$,  $m_{\rm SUSY}$ and $X_t$ are also sensitive  the precisions of fermion and vector gauge boson couplings. For the max-mixing scenario, the loop-induced $hgg$ and $h\gamma\gamma$ couplings are the main restrictions on $m_{\rm SUSY}$ when $\tan\beta>7$.

Putting all contributions together, we project the constraints onto the two-dimensional planes.  Detailed results can be found in Ref.~\cite{Li:2020glc}.   Here we only show parts of  $m_A-\tan\beta$ and $m_A - m_{\rm SUSY}$.  We also compare the sensitivity of 
indirect search via Higgs precision measurements with direct search limits at current and future LHC runs.

\label{sec:comparison}

\begin{figure}[htb]
\begin{center}
\includegraphics[width=7.5cm]{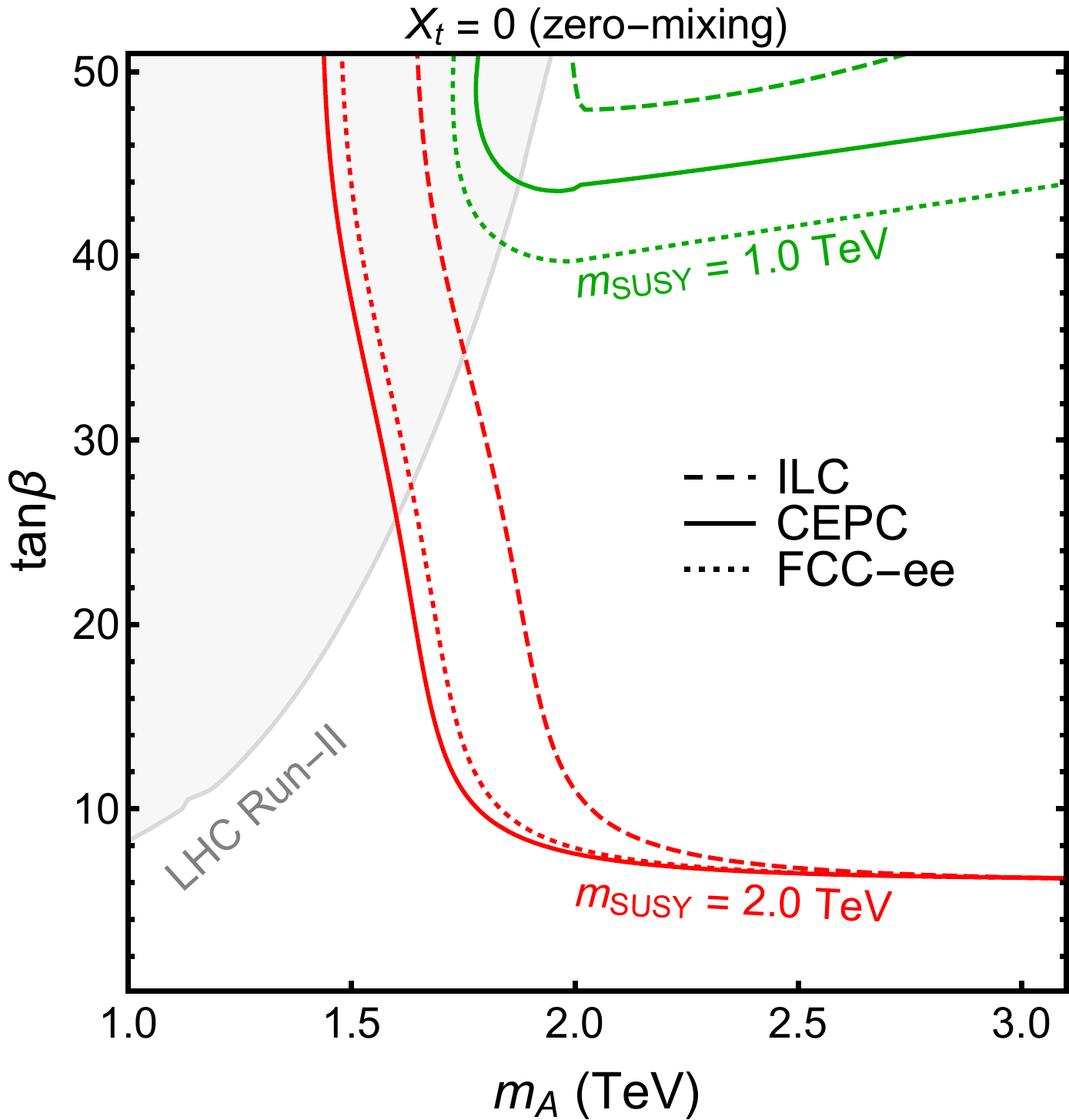}
\includegraphics[width=7.5cm]{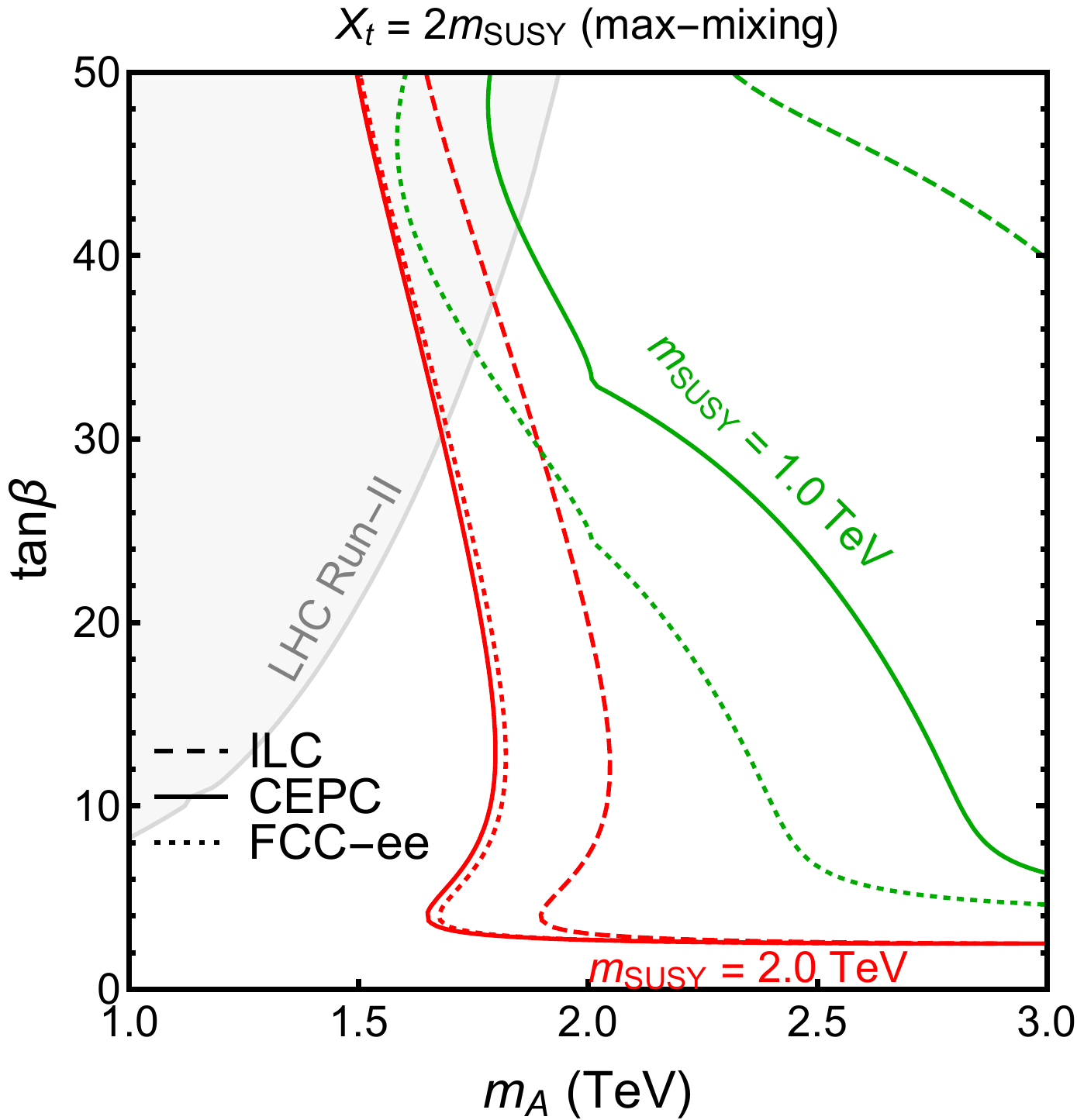}
 \caption{
 95\% C.L. allowed region in  $\tan\beta$ vs. $m_A$   for $X_t =0$ (zero-mixing, left panel)  and $X_t = 2 m_{\rm SUSY}$ (max-mixing, right panel), with  the CEPC (solid), the FCC-ee (dotted) and the ILC (dashed) precision.    The LHC Run-II direct search limits based on $A/H\to\tau\tau$~\cite{Aad:2020zxo} are shown in the grey shaded region.
 }
\label{fig:comparison}
\end{center}
\end{figure}

\begin{figure}[htb]
\begin{center}
\includegraphics[width=6.5cm]{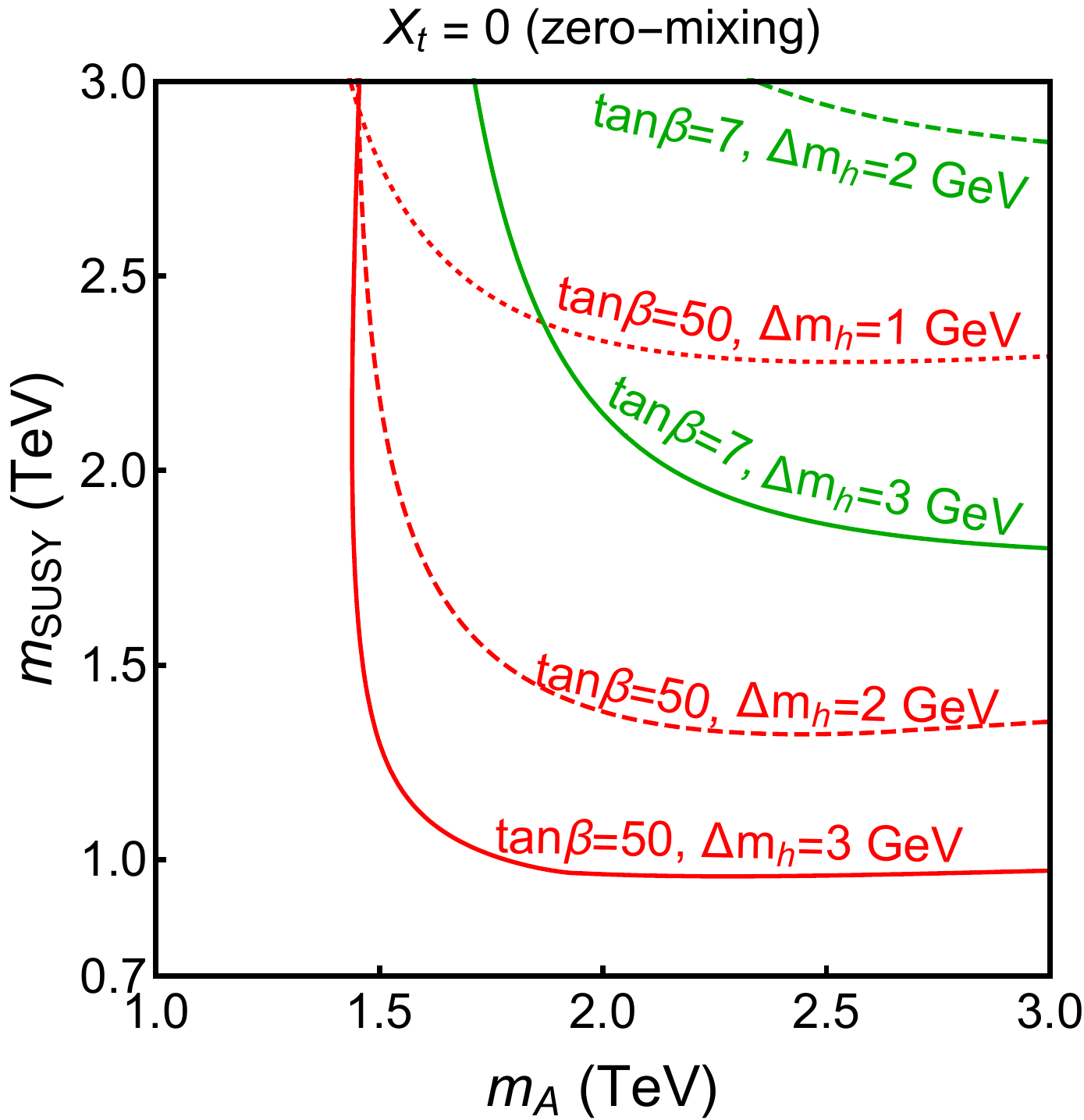}
\includegraphics[width=6.5cm]{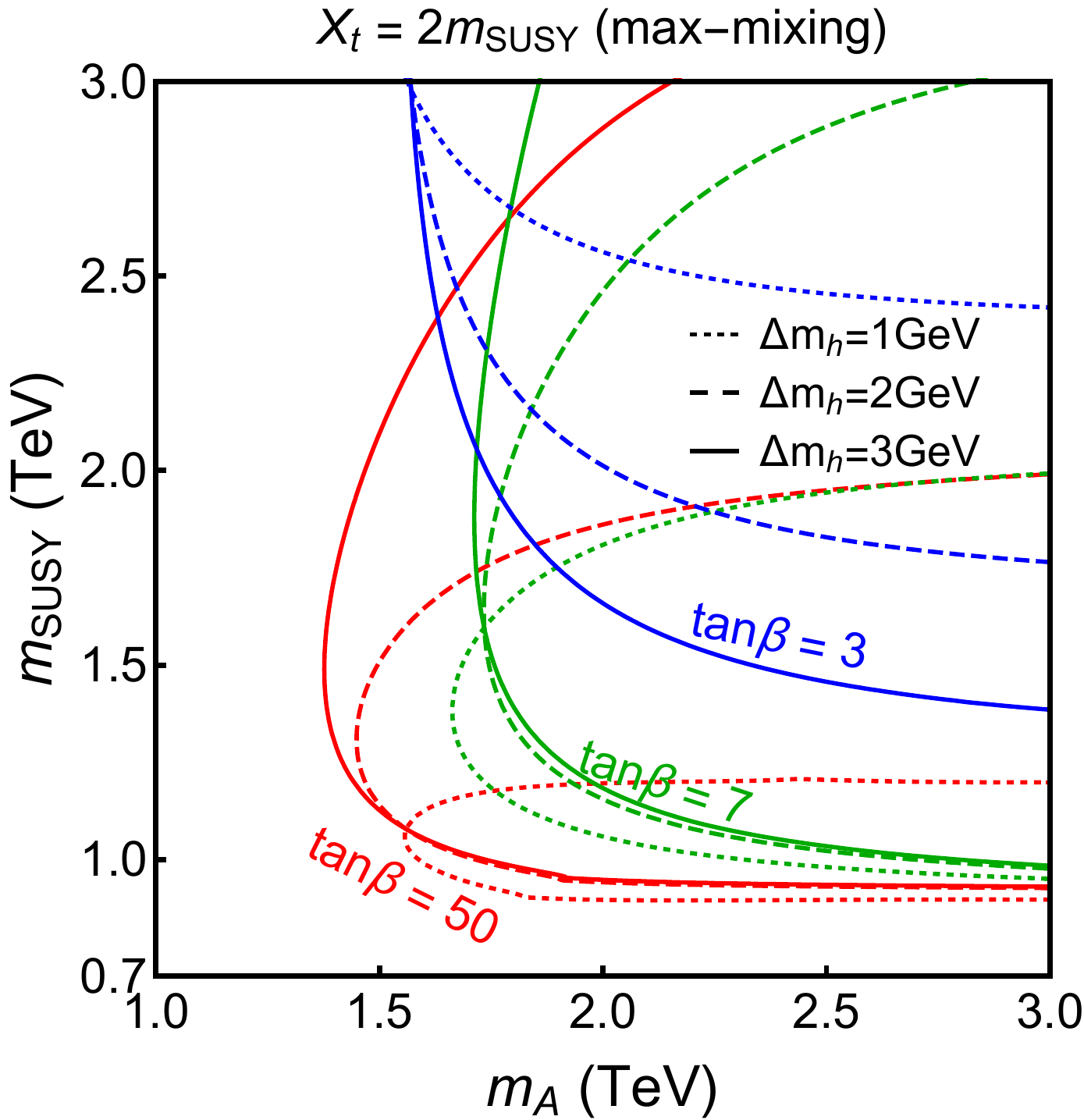}
\caption{95\% C.L. allowed region in $m_{\rm SUSY}$ vs. $m_A$ plane with CEPC precisions for $X_t =0$ (zero-mixing, left panel)  and $X_t = 2 m_{\rm SUSY}$ (max-mixing, right panel).   For each panel, different colored curve corresponds to different values of $\Delta m_h=1,\ 2, \ 3$ GeV, with region to the right of the curve allowed. 
}
\label{fig:deltamh1GeV}
\end{center}
\end{figure}

% $M_{\rm SUSY}$ vs. $X_t$, $M_A$ vs. $m_{\rm SUSY}$,

 As shown in Fig.~\ref{fig:comparison}, We obtained the 95\% C.L. allowed region given the Higgs factory precisions, and presented the result in the parameter space of  $m_A$ vs. $\tan\beta$ for various Higgs factories. In general,  $m_{\rm SUSY}<900$ GeV are excluded for both the zero-mixing and max-mixing cases.  For the zero-mixing case, when $m_{\rm SUSY}<1$ TeV, the parameter space of $\tan\beta<40$ is excluded. Limits on $\tan\beta$ get lower for larger values of $m_{\rm SUSY}$, which is sensitive in particular for $1\ {\rm TeV}<m_{\rm SUSY}<1.5$ TeV. For the max-mixing case, limits on $\tan\beta$ is much lower for $m_{\rm SUSY}=1$ TeV.    Those features   are mainly due to the Higgs mass constraint.  It also tells that FCC-ee is of stronger probe ability because of the higher luminosity.   The LHC Run-II direct search limits based on $A/H\to\tau\tau$~\cite{Aad:2020zxo} are shown in the grey shaded region, which is complementary to the indirect limits from Higgs precision measurements.
 
In Fig.~\ref{fig:deltamh1GeV}, we show the  95\% C.L. allowed region in $m_A$ versus $m_{SUSY}$ plane for $\Delta m_h=3$ GeV (solid curve) 2 GeV (dashed curve) and 1 GeV (dotted curve).      The lower limit on $m_{\rm SUSY}$ for the zero-mixing case, and the upper limit on $m_{\rm SUSY}$ for the max-mixing case   depend sensitively on the values of $\Delta m_h$.    Therefore, it is crucial to improve the precision in the $m_h$ calculation in the MSSM, which allows us to obtain tight constraints on the SUSY mass scale, in particular, on the stop sector, once Higgs precision measurements are available at future Higgs factories.

\section{Conclusion}
As a summary, we  performed a multi-variable fit to both the signal strength for various Higgs decay channels at Higgs factories and the Higgs mass.  
 We obtained the 95\% C.L. allowed region given the Higgs factory precisions, and presented the result in the parameter space of  $m_A$ vs. $m_{\rm SUSY}$, and $m_{A}$ vs. $\tan\beta$.  We found that the lower limits on $\tan\beta$ depends sensitively on the values of $m_{\rm SUSY}$ and $m_A$, in particular, for $m_{\rm SUSY}<1.5$ TeV and $m_A<2$ TeV.   Limits on $m_{\rm SUSY}$ also depend sensitively on $\Delta m_h$, indicating the importance of a precise determination of the Higgs mass in the MSSM. For $\tan\beta=50$ of the max-mixing scenario, $m_{\rm SUSY}  \in (0.8,1.2)$ TeV when $\Delta m_h= 1 $ GeV. We also compared the reach of the CEPC, the FCC-ee and the ILC.  We found that the reach of the CEPC is similar to that of the FCC-ee, while the reach of the ILC is typically better, given the slight better precision in the Higgs WBF measurements.

% \clearpage
\bibliographystyle{JHEP}
\bibliography{references}

\end{document}